\documentclass{aa501}
\usepackage{psfig}
\usepackage{epsf}
\usepackage{graphicx}

\newcommand{\Msun}{\mbox{$M_{\odot}\;$}}
\def\lsim{\;\raise0.3ex\hbox{$<$\kern-0.75em\raise-1.1ex\hbox{$\sim$}}\;}
\def\gsim{\;\raise0.3ex\hbox{$>$\kern-0.75em\raise-1.1ex\hbox{$\sim$}}\;}

\def\mathnew{\mathsurround=0pt}
\def\simov#1#2{\lower .5pt\vbox{\baselineskip0pt \lineskip-.5pt
        \ialign{$\mathnew#1\hfil##\hfil$\crcr#2\crcr\sim\crcr}}}

\def\cmc{\rm ~cm^{-3}}
\def\cms{\rm ~cm^{-2}}
\def\diff{\rm ~cm^2~s^{-1}}
\def \kms {\rm ~km~s^{-1}}
\def \cmsec {\rm ~cm~s^{-1}}
\def\ergs{\rm ~erg~s^{-1}}
\def\enf{\rm ~erg~s^{-1}~cm^{-2}}

\def\etal{{ et al. }}

\def \chan {{\it Chandra}}
\def \xmm {{\it XMM-Newton}}

\def \hcm {\hbox {\ifmmode $ atom cm$^{-2}\else atom cm$^{-2}$\fi}}
\def \arcmin {\hbox{$^\prime$}}

\def\approxgt{\mathrel{\hbox{\rlap{\lower.55ex \hbox {$\sim$}}
        \kern-.3em \raise.4ex \hbox{$>$}}}}
\def\approxlt{\mathrel{\hbox{\rlap{\lower.55ex \hbox {$\sim$}}
        \kern-.3em \raise.4ex \hbox{$<$}}}}
\begin{document}

%\thesaurus{(02.01.1; 02.18.5; 09.03.1; 09.03.2; 09.09.1: \src; 09.19.2)}

\title{X-ray Line Emission from Supernova Ejecta Fragments}

%\end{center}

\author{A. M. Bykov
}

\institute{A.F. Ioffe Institute for Physics and Technology,
          St. Petersburg, 194021, Russia}

\offprints{e-mail: byk$@$astro.ioffe.rssi.ru}

\date{Received 4 March 2002 / Accepted 26 April 2002}

\abstract{ 
We show that fast moving isolated fragments
of a supernova ejecta composed of heavy elements
should be sources of  K$_\alpha$  X-ray line emission of
the SN nuclear-processed products.
%They are also characterized by hard X-ray continuum of
%typical photon index $\leq$ 1.7.
Supersonic motion of the knots in the intercloud medium
will result in a bow-shock/knot-shock structure creation.
Fast nonthermal particles accelerated by Fermi mechanism in the
MHD collisionless shocks diffuse through a cold metallic knot, suffering
from Coulomb losses and producing the X-ray emission.
%due to radiative transitions following the removal of the 1s
%atomic electrons.
We modeled the X-ray emission from a fast moving knot of a mass
$M \sim 10^{-3} ~\Msun$, containing about $\sim$ 10$^{-4}~\Msun$ of
any metal impurities like Si, S, Ar, Ca, Fe.
The accelerated
electron distribution was simulated using the kinetic description
of charged particles
interacting with strong MHD shock. We accounted for
nonlinear effects of shock modification by the nonthermal particles pressure.
The K$_\alpha$ line emission is most prominent for the knots
propagating through dense molecular clouds.
The bow shock should be a radiative wave with a prominent infrared and
optical emission.
In that case the X-ray line spectrum of an ejecta fragment
is dominated by the 
low ionization states of the ions with the metal
line luminosities reaching $L_{\rm x} \gsim$ 10$^{31}~  \ergs$.
High resolution \xmm\ and \chan\ observations are able to
detect the K$_{\rm \alpha}$ and K$_{\rm \beta}$ line emission from the knots
at distances up to a few {\rm kpcs}.
The knots propagating through tenuous interstellar matter are of much fainter
surface brightness but
long-lived. The line spectra with higher ionization states of the ions are
expected in that case.
Compact dense knots could be
opaque for some X-ray lines and that is important for their
abundances interpretation.
The ensemble of unresolved knots of
galactic supernovae can contribute substantially to the
iron line emission observed from the Galactic Center region
and the Galactic ridge.
 }

\maketitle

\section{Introduction}
Physical mechanisms of energy release and nucleosynthesis products ejection
in supernova (SN) events are of fundamental importance for astrophysics.
Optical and UV studies of the structure of SN remnants (SNRs)
have revealed a complex metal ejecta structure with the presence
of isolated high-velocity fragments of SN ejecta
interacting with surrounding media. The most prominent manifestations
of this phenomena are the fast moving knots (FMKs) observed in some
young "oxygen-rich" SNRs in the Galaxy: Cas A
(e.g. Chevalier \& Kirshner 1979; Fesen \etal 2002),
Puppis A (Winkler \& Kirshner 1985), G292.0+1.8,
in the LMC N132D and 1E 0102.2-7219 in the SMC.

Optical FMKs in Cas A are showing very high abundances in O-burning and
Si-group elements with high Doppler velocities. They have a
broad velocity distribution around 6,000 $\kms$,
while bright radio knots are slower (e.g. Bell 1977).
The optically emitting mass of the observed knots is only about
10$^{-4} \Msun$ (Raymond 1984).

Ejecta-dominated X-ray emission from Cas A was observed with
\chan\ (Hughes \etal 2000; Hwang \etal 2001)
and \xmm\ (Bleeker \etal 2001; Willingale \etal 2002).
Willingale \etal (2002) found an emission component that is probably due to
heating of ejecta clumps sweeping up the ambient gas.
They also note that the observed Fe-K line
emission is confined to two large clumps.
Recent \chan\ images of G292.0+1.8 (Park \etal 2001)
revealed a complex structure with multiple knots and filaments on
angular scales down to the instrumental resolution.

X-ray knots were discovered with {\it ROSAT} in the Vela SNR
by Aschenbach \etal (1995) as 'shrapnels', boomerang structures outside of the main shell.
High-resolution \chan\ observations
of shrapnel A reported by Miyata \etal 2001 revealed a head-tail structure
of the apparent size $8\arcmin.4 \times 4\arcmin.1$ ($0.6 \times 0.3$ {\rm pc}
at 250 {\rm pc} distance). They estimated the gas pressure in the head to be
roughly ten times higher than that in the tail and $T_{\rm e} \sim$ 0.5 {\rm  keV}.
The oxygen abundance was 0.34$^{+0.12}_{-0.08}$ of the solar value while
that of Si was 3$^{+2}_{-1}$ times of solar.

A localized region of 6.4 {\rm  keV} emission indicating the presence of
Fe XVII or lower ionization states was found in the supernova remnant RCW 86
with {\it ASCA} instrument by Vink \etal 1997.

High resolution \chan\ observation
of the Galactic Central region reported by Wang \etal (2002)
revealed that the observed He-like Fe K$_{\rm \alpha}$ emission is largely
due to the discrete X-ray source population. Apart from the large-scale
thin plasma emission in the Galactic Center region
Bamba \etal (2002) have also observed with \chan\
many small clumps of emission lines from neutral (6.4 {\rm  keV}) to helium-like (6.7 {\rm  keV}),
with intermediate line energies between 6.5-6.7 {\rm  keV}.
The problem of the origin of the observed large scale X-ray
emission from the Galactic ridge requires a careful study of possible
classes of abundant hard X-ray sources with $L_{\rm x} \sim 10^{29} \ergs$
(e.g. Tanaka \etal 1999).
In this respect modeling of
an ensemble of X-ray point sources associated with fast moving
SN ejecta fragments seems to be necessary.

In the paper we present a model of X-ray line production
in a supersonically moving ejecta fragment. The line emission is
due to $K$-shell ionization by nonthermal particles accelerated by the
bow shock and then propagating through a cold metal-rich clump.
The optical depth effects are important in the model and they
are accounted for. Modern arcsec resolution instruments such
as that aboard \xmm\ and \chan\  are sensitive to photons up to 10 {\rm  keV}
with a good spectral resolution providing
unique possibilities to study X-ray line emission of the ejecta fragments.
These observations would allow to study stellar nucleosynthesis.

\section{Model}
Consider a fast-moving metal-rich ejecta fragment.
Supersonic motion of the fragment
(of velocity $v_{\rm k}$, radius $R$ and mass $M$) in the ambient medium
will result in a bow-shock/internal-shock structure creation
(Chevalier 1975; Sutherland \& Dopita 1995).
The bow shock has the velocity $\gsim v_{\rm k}$ and it is propagating in an
ambient medium of the standard cosmic composition. The bow shock 
creates a hot
%(of {\rm  keV} regime temperatures)
high pressure gas and also accelerates nonthermal particles.
The high pressure gas in the head of the fast moving ejecta fragment
could  drive an internal shock of velocity $v_{\rm is}$ into the fragment if $v_{\rm is}$
exceeds the sound speed in the cold matter:
\begin{equation}
v_{\rm is} \approx v_{\rm k}\, (\rho_{\rm a}/\rho_{\rm k})^{1/2} \propto v_{\rm k}\, \rho_{\rm a}^{1/2}
R^{3/2} M^{-1/2}
\end{equation}
Here $\rho_{\rm a}$ and $\rho_{\rm k}$ are the ambient gas
and the dense fragment densities, respectively. We shall also use the density
contrast $\chi = \rho_{\rm k}/\rho_{\rm a}$.
The characteristic size $R$ of 10$^{16}$-10$^{17}$  {\rm cm} is consistent with that
resolved by optical observations of Cas A
and Puppis A while large clumps were inferred by \chan\ observations
of Vela shrapnel A by Miyata \etal 2001.
For an ejecta fragment of radius $R_{16} \sim$ 3  (where
$R_{16}$ is the radius measured in 10$^{16}$  {\rm cm}) and mass
$M \sim 10^{-3}\Msun$ the density ratio
$\rho_{\rm a}/\rho_{\rm k} = \chi^{-1}$ is about 4$\cdot 10^{-4}$
for an ambient medium of number density $\sim 0.1 \cmc$.
However the ratio could be as high as $\gsim 0.1$ for the ejecta
fragments moving through a dense ambient matter of a molecular cloud.

 There is a number of factors limiting
the fast ejecta fragment lifetime.
A fast moving knot is decelerating due to the interaction with the
ambient gas.
The deceleration time of a knot can be estimated as
$\tau_{\rm d} \approx Mv_{\rm k}^2/\rho_{\rm a} v_{\rm k}^3 S \sim \chi R/v_{\rm k}$,
where $S$ is the effective crossection of the bow shock.
However, the fast knot  hydrodynamical
crushing effects could have much shorter timescale,
providing $\tau_{\rm c} \sim \chi^{1/2} R/v_{\rm k} \sim R/v_{\rm is}
\propto M^{1/2} (R \cdot \rho_{\rm a})^{-1/2} v_{\rm k}^{-1}$
(e.g. Wang \& Chevalier 2001 and references therein).
That implies $\tau_{\rm c}/ \tau_{\rm d} \sim \chi^{-1/2} \propto R^{3/2} \rho_{\rm a}^{1/2} M^{-1/2}$.
The knot ablation process with surface erosion could also be a
limiting factor.

The hydrodynamical estimation of the inner shock velocity $v_{\rm is}$
given by the Eq.(1) assumes an efficient
conversion of the bow shock ram pressure to the knot internal shock.
That is a good approximation if the effects of energy loss due to
nonthermal particle acceleration and radiative cooling of high pressure
gas behind a bow shock are negligible. The latter is the case unless the knot
is interacting with dense matter (like molecular cloud clumps).

The Alfvenic Mach number of both the bow shock and
the inner shock (if $v_{\rm is}$ is determined by the Eq.(1)) is
\begin{equation}
{\cal M}_{\rm a} = v_{\rm k} (4\pi \rho_{\rm a})^{1/2}/B \approx 460\, v_8\, n_{\rm a}^{1/2}/ \, B_{-6},
\end{equation}
where  $n_{\rm a}$ is the ionized ambient gas number density measured in $\cmc$,
$B_{-6}$ is the local magnetic field just before the shock measured in $\mu$G
and $v_8$ is the flow velocity in 10$^8~\cmsec$.
Note here that the Alfvenic Mach number of the inner shock depends on
the ambient density but on the magnetic field inside the knot.
The sonic Mach number for a shock propagating in a plasma of
the standard abundance is
\begin{equation}
{\cal M}_{\rm s} \approx 85\, v_8\, \cdot [T_4 \cdot (1 + f_{\rm ei})]^{-1/2},
\end{equation}
where $T_4$ is the plasma ion temperature measured in 10$^4$ K
and $f_{\rm ei} = T_{\rm e}/T_{\rm i}$. In general, for the processes in the precursor and
viscous velocity jump of an MHD collisionless shock wave no equilibration
between electrons and ions should be assumed (e.g. Raymond, 2001).
Both cases of $f_{\rm ei} \ll 1$ and $f_{\rm ei} > 1$ could be relevant to the
system under consideration. The photoionized gas at the radiative precursor of a
fast shock is assumed to have $T_{\rm e} \sim (1-2)\cdot 10^4$ K
and $f_{\rm ei} \gsim 1$ (e.g. Shull \& McKee 1979). On the other hand,
plasma heating due to MHD waves dissipation in the vicinity of the
viscous subshock heats mostly the ions, providing $f_{\rm ei} < 1$.

\subsection{The effect of nonthermal particles on a strong bow shock}

For strong collisionless shocks in a magnetized plasma the nonthermal particle
acceleration effect is expected to be efficient and a significant fraction of
ram pressure is transferred to high energy particles
(ions for nonrelativistic shocks).
The shock transition of a strong shock of the total Mach number ${\cal M}{\rm t} \gg$ 1
is broadened because of the upstream gas deceleration by nonthermal
particle pressure gradient ahead of the viscous gas subshock of a
modest Mach number ${\cal M}_{\rm sub} \leq$ 3
(see for a review Blandford \& Eichler 1987; Jones \& Ellison 1991;
Malkov \& Drury 2001).

The total compression ratio $R_{\rm t}$ of a strong MHD shock modified by an
efficient nonthermal particle acceleration  depends on
the energy flux carried out by escaping
nonthermal particles  and the effective adiabatic
exponent (e.g. Malkov \& Drury 2001). 
On the other hand the distribution function of nonthermal particles
and the bulk flow
profile in the shock upstream region are sensitive to the total compression
ratio $R_{\rm t}$. Thus, the exact calculation of the escape flux can be
performed only in fully nonlinear kinetic simulations. Nevertheless, some
approximate iterative approach can be used to make
the distribution function consistent with the shock compression.
The subshock is the standard gas viscous shock with the compression  ratio:
$R_{\rm sub} = (\gamma_{\rm g} +1) {\cal M}_{\rm sub}^2$/
$[(\gamma_{\rm g} -1){\cal M}_{\rm sub}^2 + 2]$.  The downstream ion
temperature $T^{(2)}$ behind the modified shock structure can be calculated
if $R_{\rm t}$ and $R_{\rm sub}$ are known for the shock of the given velocity and
it could be very different from that obtained with
the standard single-fluid Rankine-Hugoniot law.
The electron temperature just downstream of the shock is expected to
be lower than $T^{(2)}$.

An exact modeling of the collisionless shock structure with nonthermal particle
acceleration effect requires a nonperturbative selfconsistent description
of a multi-component and multi-scale system including strong MHD turbulence
dynamics. That modeling is not feasible at the moment. We shall use some
simplified description of a strong shock with some apropriate
parameterization of governing parameters which are:
$(a)$ the nonthermal particle diffusion coefficients;
$(b)$ the ion injection rate;
$(c)$ the maximum momentum of accelerated particles. The
gas heating mechanism due to MHD waves dissipation in the shock precursor
must also be specified in that simplified approach, providing the connection
between the total compression and the gas subshock Mach number
${\cal M}_{\rm sub}$.
Assuming that the main heating mechanism of the gas in the precursor
region is due to Alfven waves dissipation, Berezhko \& Ellison (1999)
obtained a simple relation for the total compression of the shock
$R_{\rm t} \approx 1.5 {\cal M}_{\rm a}^{3/8}$, while the subshock had
${\cal M}_{\rm sub} \sim 3$. The relation is valid under the condition
of ${\cal M}_{\rm s}^2 \gg {\cal M}_{\rm a}$ in the far upstream flow.
They also noted that the subshock compression
ratio is not too sensitive to the ion injection rate if the rate
exceeds $5 \cdot 10^{-5}$ to provide the shock
modification by the accelerated ions.
We obtain then the ion temperature just behind the modified strong shock
 (measured in 10$^6$ K)
\begin{equation}
T^{(2)}_6 \approx 0.32 \cdot \phi({\cal M}_{\rm sub}) \cdot v_{\rm k8}^{5/4} n_{\rm a}^{-3/8} B_{-6}^{3/4}(1 + f_{\rm ei})^{-1},
\end{equation}
where $\phi({\cal M}_{\rm sub}) = [2 \gamma_{\rm g} {\cal M}_{\rm sub}^2 - (\gamma_{\rm g} -1)]
 /[(\gamma_{\rm g} -1){\cal M}_{\rm sub}^2 + 2]$,
 and $\phi(3) \approx 3.7$ for $\gamma_{\rm g}$= 5/3.
Then the electron and ion temperatures equilibrate in the postshock region.
The initial electron temperature in the shock downstream depends also
on the collisionless electron heating
(see e.g. Bykov \& Uvarov, 1999; Laming 2001).
The turbulent gas heating due to acoustic instability wave
dissipation could be more efficient however, if the acoustic instability
develops (Malkov \& Drury 2001).

The maximal momenta of
the accelerated ions are limited by MHD wave damping due to ion-neutral
collisions in the case of a fragment interacting with dense molecular
cloud. The effect was accounted for following the work by
Drury \etal (1996).
The maximal momenta of the ions are $\lsim 10^2$  GeV/c in
that case. In the case of a fragment propagating through tenuous medium
the maximal momentum is limited by the finite scale of strong scattering
region with typical values of the maximal momentum $\gg 10^2$  GeV/c.

It should be noted that we still have no direct observational evidences
for the cosmic ray
ions accelerated in SNR shocks. At the same time  there are direct evidences
for ion and electron acceleration by strong cosmic-ray modified shocks
in the interplanetary medium (e.g. Terasawa \etal 1999).
In the interplanetary shock of velocity $\sim$ 375 $\kms$
(in the solar wind frame) studied by Terasawa \etal (1999) with
the {\it GEOTAIL} satellite the partial pressure of accelerated nonthermal
particles was found to be in a rough balance with the thermal proton pressure
in the shock downstream. That implies a high efficiency of nonthermal
ion production. The diffusion coefficients for both electrons and ions
were estimated to be $\sim$ 10$^{18}~\diff$ in the shock vicinity
that is at least two order of magnitude smaller than
the typical interplanetary values (Shimada \etal 1999).
For SNR shocks Dorfi (2000) estimated the
ion acceleration efficiency to be 0.24 for some reasonable parameters.

The observed synchrotron
emission from SN shells is considered for a long time as a direct evidence
for relativistic electron acceleration. The efficiency of transformation
of the kinetic energy of a nonrelativistic shock bulk flow into the
nonthermal electron population is not expected to be very high, being
below a few percent. However even a percent range efficiency of the
nonthermal electron acceleration would provide an observable effect
of K$_{\rm \alpha}$ photon emission from metallic knots irradiated by
high energy electrons.
Bulk heating and ionization  of the knot interior due to
deep penetration of energetic ions and electrons could also be important
for enhanced ablation.

\subsection{Nonthermal electron acceleration}
To simulate the spectra of nonthermal electrons
in the fragment body  we solve a kinetic equation for
the distribution function of the electrons, accounting for injection,
diffusive transport, advection,
first and second order Fermi acceleration and synchrotron and
the Coulomb losses.
Electron kinetics in supercritical collisionless shocks
was modeled by Bykov \& Uvarov (1999).
They showed that strong MHD fluctuations generated by kinetic
instabilities of ions are responsible for  heating and pre-acceleration of
nonthermal electrons on a very fine scale, of the order of several hundreds
of  inertial lengths, in the vicinity of the viscous jump (subshock) of a
collisionless shock.
We have described the details of the code, diffusion model
and loss functions used elsewhere
(Bykov \& Uvarov 1999; Bykov \etal 2000).

 There are three zones in the one-dimensional model: the pre-shock region (I),
the shock transition region (II), and the post-shock flow (III),
from where nonthermal emission from shock-accelerated
particles originates. In order to calculate the spectra of nonthermal
electrons in these regions, we used the following kinetic equation for the
nearly-isotropic distribution function $N_i(z,p)$ ($i$ = I -- III):
\begin{eqnarray}
    &&
           k_i(p) \: \frac{ \partial^2 N_i(z,p) }{ \partial z^2 } +
      \frac{1}{p^2} \: \frac{\partial}{\partial p} \: p^2 [D_i(p) \:
      \frac{\partial N_i}{\partial p} + L_i(p) \:N_i]  -     \nonumber \\
    &&
   -   u_i(z)\: \frac{ \partial}{ \partial z }\: N_i +
     \frac{p}{3} \:\frac{\partial}{\partial p}\: N_i\:
     \left[\frac{ \partial}{ \partial z }\: u_i \right]  = 0 .
\end{eqnarray}

This Fokker-Planck-type equation takes into account diffusion and
advection [bulk velocity $u_i(z)$] of electrons in phase space
due to interactions with MHD waves and the large-scale MHD flow.
Here $z$ is the coordinate along the shock normal.
$L_i(p)$ is the momentum loss rate of an electron
due to Coulomb collisions in a partially ionized plasma and
synchrotron/inverse Compton radiation.
The momentum diffusion coefficient $D(p)$ is responsible for
the second order Fermi acceleration,
and $k_i(p)$ is the fast particle spatial diffusion coefficient.
For low-energy electrons the
Coulomb and ionization losses
are important everywhere except for the narrow shock
transition region (II), where  acceleration is fast enough to overcome
the losses and where nonthermal electron injection occurs.

To accelerate the electrons injected in the shock transition region
to relativistic energies  MHD turbulence should
fill the acceleration region. As we know from the interplanetary 
shock observations, turbulence generation is a generic property
of collisionless MHD shocks.  We used the  parameterization of the
momentum dependent diffusion coefficient  $k_i(p)$
of an electron of momentum $p$, described in details
by Bykov \etal (2000). The diffusion
coefficients are momentum dependent with $k_i(p) \propto k_0 \cdot v p$
above 1 {\rm  keV}. The normalization value $k_0$ is the
electron diffusion coefficient at 1 {\rm  keV}.

The electron temperature in the (photo)ionized shock precursor
was fixed to $T_{\rm e} = 2 \cdot 10^4$ K for the fragments propagating
in the dense clouds of the gas of number density $\gg$ 1 $\cmc$.
The fragment
velocity should exceed $\sim 250 \kms$ to provide enough UV radiation
to complete H and He (to He$^+$) ionization in the precursor
of the bow shock, if it is modified by the nonthermal particle pressure.
The velocity is higher than the value of $\sim 110 \kms$ derived
by Shull \& Mckee (1979), because of the weaker gas heating by the
modified MHD shock structure with a smooth precursor
and a gas subshock of ${\cal M}_{\rm sub} \sim 3$.
It should be noted however that even at lower shock velocities
where the upstream photoionization is inefficient, the preshock ionization
could be provided by the fast particles accelerated at the collisionless
shock.
For the fragments propagating in hot tenuous plasma the electron
temperature was assumed to be equal to the plasma temperature
(i.e. $f_{\rm ei}$ = 1).

The energy spectrum of the nonthermal electrons in the bow shock
of a metal-rich ejecta fragment is shaped by the joint action of the
first and second order Fermi acceleration in a turbulent plasma with
substantial Coulomb losses. 
The electron spectra are calculated for different values of
the bow shock velocity assuming that the diffusion coefficient is
parameterized as in Bykov \etal (2000) with different values of
$k_0$ for 1 {\rm  keV} electron for different environments.
The minimal value of the diffusion coefficient
of an electron is $k_0^{min} \approx 6.6\cdot 10^{16} B^{-1}_{\mu G} (E/{\rm  keV}) \diff$.
Note that the minimal value is only achievable if strong fluctuations of
magnetic field are present at the electron gyro-radii scale. For
slightly superthermal electrons the fluctuations cannot be generated
by the ions. Whistler type fluctuations could provide the required
electron scale fluctuations (e.g. Levinson 1996). However in the case
of a strong MHD turbulence the electron
transport by the vortex type fluctuations of ion gyro-radii scale
would provide the effective diffusion coefficient much larger than
$k_0^{min}$ (Bykov \& Uvarov, 1999).
We consider here rather a conservative case of that large
diffusion coefficient.

The electron acceleration model described above is one-dimensional.
The transverse gas flow of velocity $u_{\perp}$
behind the bow shock front would affect
the nonthermal particle distribution. The exact 2-D and 3-D modeling
of the postshock flow with account of shock modification effect is not
feasible at the moment.
However, the transverse flow effect on high energy electrons is only marginal
if the particle diffusion time across the transverse flow
is shorter than the transverse advection time scale i.e.
$\delta_{\perp}/R  \cdot u_{\perp} \delta_{\perp}/k(p) <$ 1.
Here $\delta_{\perp}$ is the width
of the transverse gas flow behind the bow shock 
(typically it is a relatively small fraction of the fragment radius R). 
In our case the condition is
satisfied for 100 {\rm  keV} - {\rm MeV} regime electrons dominating the $K_\alpha$ lines
production in the ejecta fragment body. We account for the presence of a
postshock cooling layer of the shocked ambient matter situated
in front of the metal rich fragment.
The curves 1-3 in Fig. 1 illustrate the effect of Coulomb losses
on low energy nonthermal electrons diffusing through the postshock layer.

%Fig1*******************************

%%%%%%%%%%%%%%%%%%%%%%%%%%%%%%%%%%%%%%%%%%%%%%%%%%%%%%%%%% Fig 1
\begin{figure}
\centering
\epsfxsize=76mm
\epsffile[70 60 440 700 ]{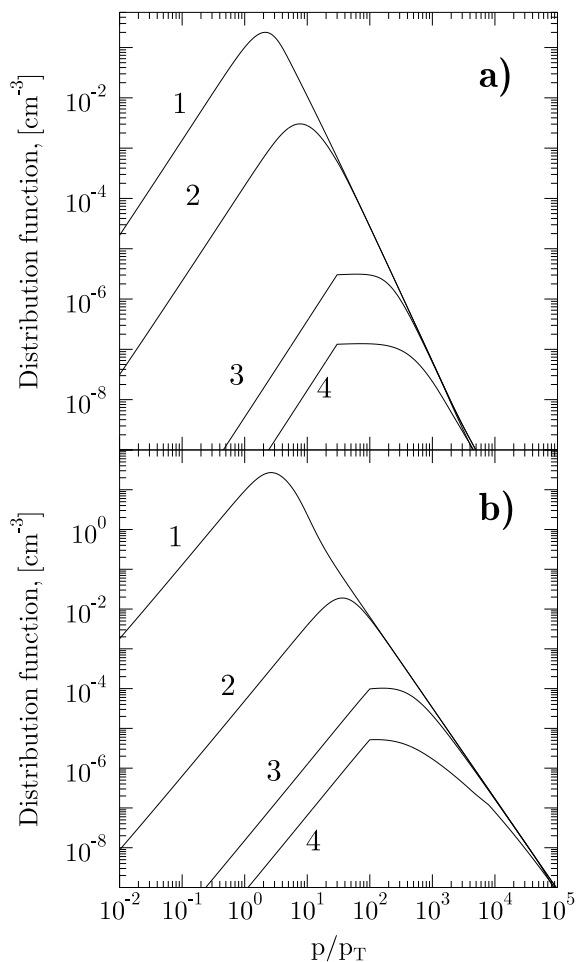}
\caption{
The nonthermal electron distribution function
(${\tilde N(p,x)} = N(p,x)p^2p_{\rm T}$) as a function of dimensionless
particle momentum $p/p_{\rm T}$
($p_{\rm T}$ is the momentum of a thermal electron in the upstream region).
We show the distributions at different
distances from the bow shock plane for different ambient media:
(a) a fragment of velocity $v_{\rm k} = 3,200 \kms$
in a tenuous medium;
(b) an SN fragment of velocity
$v_{\rm k} = 1,080 \kms$ in a molecular cloud (see Section 2.3 for details).
The depths from the bow shock plane are 10$^6$, 10$^{12}$, 10$^{15}$,
3$\cdot$ 10$^{16}$  {\rm cm} for the curves 1--4, respectively.
}
\label{fig1}
\end{figure}
%%%%%%%%%%%%%%%%%%%%%%%%%%%%%%%%%%%%%%%%%%%%%%%%%%%%%%%%%%%%%%

\subsection{Ambient conditions and the fragment state}
As a generic example we consider a fast moving fragment
of $v_{\rm k} \sim$ 1,000 $\kms$,
 mass $M = 10^{-3} ~\Msun$ and radius R$_{16}$ = 3 dominated by oxygen
 and containing about $\sim$ 10$^{-4}~\Msun$ of any metal impurities like
Si, S, Ar, Ca, Fe. That fragment we refer as "the standard knot".

The electron distribution was simulated here for two distinct representative
cases of the ambient environment. The first one is an SN fragment
propagating through dense ambient medium of number density $\sim$ 10$^3~\cmc$,
 relevant to the SN explosion in the vicinity of a molecular cloud.
If the ambient magnetic field follows the scaling
$B_{\rm a} \propto n^{1/2}$ (e.g. Hollenbach \& McKee 1989) one may expect
the magnetic field values of $\gsim$ 100 $\mu$G in the dense ambient
medium (c.f. Blitz \etal 1993). The second case is a
SN fragment propagating through   tenuous medium
of number density
$\sim$ 0.1$~\cmc$, temperature T $\gsim$ 10$^4$ K and magnetic
field of $\mu$G range.
Correspondingly, the diffusion coefficient normalization factors
for {\rm  keV} regime electrons
were $k_0 \sim 10^{18} \diff$ in the dense medium case and
$k_0 \lsim 10^{20} \diff$ for the tenuous medium.
The gas ionization state in the bow shock downstream was simulated
here in the way like that described
in the paper of Bykov \etal (2000).

The gas ionization state inside an oxygen rich fragment body
was estimated following the model by Borkowski \& Shull (1990).
They studied the structure of a steady state
radiative shock (inner shock) of velocities $v_{\rm is}$
from 100 to 170 $\kms$  (c.f. Eq.(1)) in a pure oxygen gas
with electron thermal conduction.
The postshock temperature
behind the 140 $\kms$ shock with thermal conduction was found to be
270,000 K in the immediately postshock region and dropped to roughly
4,700 K after the oxygen depth about $2 \cdot 10^{15} \cms$ in
the downstream photoionization zone. The dominant ionization state of oxygen
in the shock radiative precursor and in the immediate postshock region
was $OIII$ while it was $OII$ in the downstream photoionization zone.
Nonthermal electrons could also be an important factor to model
the dominant ionization state of the ions in the fragment body
(Porquet \etal 2001).

In Fig. 1 the nonthermal parts of the electron
distribution functions ${\tilde N(p,x)} = N(p,x)p^2p_{\rm T}$ are shown
at the different depths of an SN fragment.
Here $p_{\rm T}$ is the momentum of a thermal electron in the far upstream
region. Note that the upstream gas electron temperatures
are different for the two ambient media we displayed in the Fig. 1.
In Fig. 1 we showed only nonthermal parts of the electron distributions
because the electron temperatures in both presented cases are below those
required for efficient $K_\alpha$ lines production in the ejecta fragments.
The threshold for K-shell ionization of Si is 1.8 {\rm  keV}
 (i.e. $p/p_{\rm T} >$ 33 in Fig. 1)
while that for Fe is about 7.1 {\rm  keV}  (i.e. $p/p_{\rm T} >$ 64 in Fig. 1).
Fast electrons accelerated in the
MHD collisionless shock diffuse
through the postshock layer and cold metallic knot suffering the
Coulomb losses as it is clear from the Fig.1.
They are producing the $K_\alpha$ lines in the fragment body
due to radiative transitions following the removal of the 1s
atomic electrons.

\subsection{K-shell ionization and X-ray line emissivity}
To calculate the production rate of characteristic X-ray line
emission from the metal rich fragment irradiated by an intense flux
of energetic electrons accelerated by an MHD shock wave
the atomic inner shell ionization cross sections and
the fluorescent yields are required. The experimental electron-impact
$K$-shell ionization
crossections and the fluorescent yields were compiled recently
by Liu \etal (2000).
The standard Bethe theory of inner shell
ionization (e.g. Powell 1976) was used to fit the data for
nonrelativistic energies of ionizing electrons,
from the threshold up to 50 {\rm  keV} and the relativistic formulae by
Scofield (1978) above 50 {\rm  keV}.

We fit the K-shell ionization cross section following
the parameterization given by  Scofield (1978)
\begin{equation}
\sigma_i = \frac{A}{\beta^2} \cdot (b + b_1 + b_2/\epsilon + b_3/p^2 + b_4\cdot b/p + b_5/p^4)
\end{equation}
Here the crossection $\sigma_i$ is given as a function
of the incident electron momentum $p$
(measured in $m_e c$), $b$ = ln$(p^2) - \beta^2$, $\beta = v/c$. The fitting
parameters $A,~b_1 - b_5$ were calculated by Scofield (1978) for the ions of
$Z \geq$ 18. We corrected the fitting parameters to account for the present
laboratory measurements compiled by Liu \etal (2000).

The logarithmic asymptotic of the crossection at relativistic energies,
given by the Eq.(6)  is important
for our modeling.
One can see in the Fig.1 (curves 3 and 4) that the electron
 spectra have rather flat maxima at mildly
relativistic energies ($p/p_{\rm T} \gsim 400$)
in the most of the volume of a dense metal-rich fragment.
Thus, the K-shell ionizations
by relativistic electrons are substantial.

The radiative decay of K-shell vacancy induced by fast electron results
in $K_{\rm \alpha}$ $(2p \rightarrow 1s)$ and
$K_{\rm \beta}$ $(3p \rightarrow 1s)$ lines production.
The $I(K_{\rm \beta})/I(K_{\rm \alpha})$ intensity ratio is about 0.14
(e.g. Scofield 1974) for
FeI-FeIX and it is decreasing for higher iron ionization stages
(e.g. Jakobs \& Rozsnyai 1986). The ratio is an increasing function of
the atomic number $Z$. Note however, that the
optical depths for $K_{\rm \beta}$ and $K_{\rm \alpha}$ lines could be different,
especially for highly ionized iron, and that would affect the observed line
ratio.
Thus, observing the ratio $I(K_{\rm \beta})/I(K_{\rm \alpha})$ one may constrain
the source  optical depth. Both $I(K_{\rm \alpha})$ and $I(K_{\rm \beta})$ Fe
lines were detected recently
with {\it Chandra} from Sgr B2 giant molecular cloud
by Murakami \etal (2001). The observed ratio is somewhat
below 0.2 indicating some possible optical depth effect,
but it is still marginally consistent with the value 0.14 expected for
a transparent system.

It is important to note here that shock-accelerated
energetic nuclei can also provide
efficient K-shell ionization.
The emission spectra from the decay of
inner-shell ionizations produced by a
collision with an {\rm MeV}-regime ion have multiple satellites (peaks)
at energies higher than that of the electron-induced $K_{\rm \alpha}$ line
(e.g. Garcia \etal 1973).
We shall discuss the effect of ions in detail elsewhere.

\subsection{The optical depth effect}
The maximal column density of a metal (of atomic weight $A$) in a spherical 
ejecta fragment $N^{(A)}_{\rm max} = 5.6\cdot 10^{22}~(M^{(A)}/\Msun)~A^{-1}~R_{17}^{-2}~~ {\rm cm}^{-2}$
can be high enough to provide optical depths $\tau \gsim 1$.
We use below $N^{(A)} = (1/2) N^{(A)}_{\rm max}$
as the resonant scattering column
density in the model assuming the shock situated at the fragment center.
The resonant line scattering effect (see, for the review, Mewe 1990) can be
important for the metal rich SN fragments.
The optical depth at the line center of an ion 
due to the resonant line scattering was given by Kaastra \& Mewe (1995)
and can be expressed as:

\begin{equation}
\tau \approx 1.34~f_{\rm abs}~N_{15}^{(A)} E_{\rm ph}^{-1} (A/T)^{1/2}
(1 + 0.5 w_6^2 A/T)^{-1/2}.
\end{equation}
where $E_{\rm ph}$ (in {\rm  keV}) is the line photon energy, $f_{\rm abs}$ is the
absorption oscillator strength ,$T$ is the ion temperature
(in eV), $N_{15}^{(A)}$ is  the ion column density
(in 10$^{15}$  {\rm cm}$^{-2}$), $w_6$ is the
gas micro-turbulence velocity (in 10 $\kms$).

To describe the radiative transfer for the line radiation
produced inside the absorbing medium we used a simplified
escape probability approach. The mean escape probability 
$p_{\rm f}(\tau)$ can be
approximated as $p_{\rm f}(\tau) \approx [1 + a \tau]^{-1}$,
where $a \approx 0.43$ for $\tau \lsim 50$ (Kaastra \& Mewe 1995).
The photon absorption due to the resonant scattering of K-shell line is most
important for the ions with an incomplete L-shell
(e.g. OI, Si VI, Ar X, Fe XVIII) and relatively low fluorescent yields.
The resonant absorption is not effective for the neutral
or the low ionization stages of Fe, Ca, Ar, Si with completed L-shells
where we applied the photoabsorption by
Morrison \& McCammon (1983).

\section{X-ray emission from an SN fragment interacting with a molecular cloud}

Core-collapsed SNe from massive progenitors are expected to be
correlated with massive molecular clouds. The most
spectacular manifestation of the phenomena are the SNRs in starburst galaxies
like NGC 253, Arp 220, M82 etc. (see e.g. Chevalier \& Fransson 2001).
There are also star-forming molecular clouds Sgr A, Sgr B2
in the Galactic Center region with
a strong CS emission from dense molecular gas
(e.g. Blitz \etal 1993). To model such a case we considered a SN
fragment of radius 3$\cdot$10$^{16}$  {\rm cm} and of
oxygen mass of 10$^{-3}~\Msun$, containing also
$\sim$ 10$^{-4}~\Msun$ of an impurity (Fe,Ar,Si). The fragment is
propagating through
a molecular clump of a number density 10$^{3}~\cmc$ with a velocity
$\gsim$ 1,000 $\kms$.
The magnetic field value in the cloud is about 100 $\mu G$
and $k_0 \approx 10^{18} \diff$ at 1 {\rm  keV}.

%Tab1****************

\begin{table}
\caption{K-shell line luminosities
of the fragment interacting with a molecular cloud }
\label{lines}
\medskip
\centering
\begin{minipage}{8cm}
\begin{tabular}{|c||c|c|c||c|c|}
\hline
 Line\footnote{The luminosities are in 10$^{38}$ ph~s$^{-1}$}
& \multicolumn{3}{c}{ $ v_{\rm k}(~\kms)$}
                       & $\tau_{\rm max}$\\
     & 1,080  & 1,620  & 2,700 & \\
\hline
\hline
O (0.54 {\rm  keV}) & 40.4 & 104 & 1638 & 33,880\\
\hline
Si (1.7 {\rm  keV}) & 1.3 & 10 & 48 & 592\footnote{The absorption depths
can be applied only for the ionization states Si VI, Ar X, Fe XVIII and 
higher.}\\
\hline
Ar (2.9 {\rm  keV}) & 0.5 & 4  & 20 & 272$^b$\\
\hline 
Fe (6.4 {\rm  keV}) & 0.4 & 3 & 15 & 78 $^b$\\
\hline
  $T^{(2)}$ [10$^7$ K]  & 0.3 & 0.5 & 1.0 &\\
\hline
\hline
\end{tabular}
\end{minipage}
\end{table}

%Tab ******************************

We calculated the local emissivities
of K-shell lines through the fragment depth
as well as the integrated
line luminosities.
Table 1 contains the luminosities
of K$_{\rm \alpha}$ lines of O, Si, Ar, Fe.
These luminosities are not corrected for the optical depth effect.
We have also given the maximal depths $\tau_{\rm max}$ for resonant scattering
calculated under the assumption that the column density of a given
charge state
of the ion A is equal to $N^{(A)}_{\rm max}$. To obtain a real
estimation one should correct the depth for the actual column densities
of the ions with incomplete L-shells.
Note that in Tables 1 and 2 we presented only
the integral luminosities  of K$_{\rm \alpha}$ complex.
A prediction for the line shape depends on the details of the
ion charge state profile inside the fragment. The exact modeling of the
ionization profile is beyond the scope of the present paper.
Nevertheless, one could see that in the case
when the bow shock is radiative (see below)
the dominant iron charge state in the fragment
becomes lower than Fe XVIII, providing the low resonant absorption depth and
the line centroid to be close to 6.4 {\rm  keV}.
We assumed in the Table 1 that the
gas micro-turbulence velocity  $w_6 = 1$,
and the ion temperature $T \lsim 10^4$K  in the fragment body.
The ion temperatures T$^{(2)}$ presented in the Tables 1 and 2 were calculated
for the position just after the viscous subshock. The observable electron
temperatures in the  postshock relaxation  region are somewhat lower.

As seen in Table 1, an increase in
the fragment velocity results in a strong enhancement of
the line luminosities.
For the lower knot velocities the line emission
drops down drastically because the Coulomb losses are dominating in that case.
Note however that if the value of $k_0$ was less than
10$^{18} \diff$ (at 1 {\rm  keV}) even the fragments of
lower velocity ($v_{\rm k} < 1,000 \kms$) could provide a substantial line
luminosity. A fast fragment interacting
with CS-emitting gas of density $\gsim$ 10$^4 \cmc$ could reach
even higher line luminosity if $k_0 \sim 10^{17} \diff$ (at 1 {\rm  keV}).
Since the larger fragments  (for a given knot mass) are
more transparent for the K-shell lines of oxygen (and other elements)
a fast SN fragment of $R \gsim 3\cdot 10^{17}$  {\rm cm}, propagating
through the inter-clump medium of density $\gsim 10 \cmc$, might have
a prominent oxygen K-line.

A fast fragment of a larger scale
$\gsim 10^{17}$  {\rm cm},  of the same mass 10$^{-3}~\Msun$,
entering the molecular clump will have $\rho_{\rm a}/\rho_{\rm k} \gsim 1$ and
 would drive a strong shock into the metal-rich fragment.
Such a fragment should be a source of gamma-ray lines and
also light (and other spallogenic origin) elements produced by accelerated 
ion interactions
with the metal-rich knot. It would appear as a bright transient source.

\section{X-ray emission from an SN fragment in a low density medium}

Many of the isolated SN fragments are propagating through a low density
hot environment. That concerns both a relatively low velocity fragments
moving inside the forward shock radius of SN and fast velocity fragments
of an SN exploding in a low-density environment of the number density
$n \approx 0.1 \cmc$, gas temperature T $\approx 2 \cdot 10^4$ K and
the magnetic field value $\sim$  3 $\mu G$.
We first considered the same SN fragment
as that in the dense medium, but in a wider velocity range
$1,000 < v_{\rm k} < 7,000 \kms$  because
even very high velocity fragments are long-lived in the low density
environment. The diffusion coefficient normalization at 1 {\rm  keV}
was fixed to be $k_0 \approx 3\cdot 10^{19} \diff$.
We summarize the simulated X-ray line luminosities
(measured in 10$^{36}$ photon s$^{-1}$) in Table 2.
One can see that the X-ray line luminosities
$L_{\rm x} <$ 10$^{29}~ \ergs$ (per 10$^{-4}~\Msun$ of Si, S, Ar, Ca, Fe)
are predicted from an individual SN fragment of the scale
3$\cdot$10$^{16}$  {\rm cm} in a tenuous medium.
In Table 2 we assumed  that the
gas micro-turbulence velocity  $w_6 = 1$,
and the ion temperature $T \lsim 10^4$K  in the fragment body.

However, simulations of somewhat larger $\gsim$ 10$^{17}$  {\rm cm} fragments
show that the luminosity corrected for the absorption is increasing
to $L_{\rm x} \gsim$ 10$^{30}~ \ergs$ and even higher due to decreasing
of the optical depth and Coulomb losses.
It is important that the large scale fragments are much thinner
(and hotter) providing a substantial amount of ions
in high ionization states. These faint transparent fragments would contribute
substantially to the observed diffuse X-ray line emission of highly
ionized matter.
Note that the fragment deceleration time is $\propto M n_{\rm a}^{-1} R^{-2}$.
 That implies that the fragments
in the old remnants could only be observed if they spent most of the
time in the tenuous medium.

\subsection{A model of the shrapnel A in the Vela SNR}

We simulated  the line emission from
a fragment of a scale $\sim$ 10$^{18}$  {\rm cm} to model the Vela shrapnel A
discovered by Aschenbach \etal (1995) and recently studied with \chan\ by Miyata \etal (2001).
The oxygen-dominated fragment of mass $M \sim 10^{-2}\Msun$ 
and velocity
$v_{\rm k} \gsim 10^8 ~\cmsec$ would have the deceleration time about
10,000 years in an ambient medium of $n \approx 0.1 \cmc$.
We found that the temperature behind the fragment bow shock dominated by
nonthermal particles is about 0.5 {\rm  keV}.
The silicon line at 1.8 {\rm  keV} would have the luminosity
$L_{\rm x} \sim$ 10$^{30}~ \ergs$ if $\sim 10^{-3}\Msun$ of Si is contained
in the fragment and the oxygen line at 0.6 {\rm  keV} --
$L_{\rm x} \gsim$ 10$^{31}~ \ergs$.
The resonant absorption depth
of the Si K-shell line is $\tau_{\rm max} \sim 0.5$   while that of oxygen is
$\tau_{\rm max} \sim 30$ assuming the
gas micro-turbulence velocity  $w_6 = 10$,
and the ion temperature $T \gsim 10^6$K  in the large diluted fragment.
The mean escape probabilities $p_{\rm f}(\tau)$ are about 0.8 for Si and
0.07 for oxygen. The optical depth effect could account for
the apparent Si overabundance observed by Miyata \etal (2001).

%Tab2****************

\begin{table}
\caption{K-shell line from the fragment interacting with low-density gas}
\label{lines}
\medskip
\centering
\begin{minipage}{8cm}
\begin{tabular}{|c||c|c|c||c|c|}
\hline
 Line\footnote{The luminosities are in 10$^{36}$ ph~s$^{-1}$}
& \multicolumn{3}{c}{ $ v_{\rm k}(~\kms)$}
                       & $\tau_{\rm max}$\\
     & 1,600  & 3,200  & 6,400 & \\
\hline
\hline
O (0.5-0.6 {\rm  keV}) & 38.0 & 66.5 & 99.8 & 33,880 \\
\hline
Si (1.7-1.8 {\rm  keV}) & 2.6 & 4.5 & 6.7 & 592\footnote{The absorption depths
can be applied only for the ionization states Si VI, Ar X, Fe XVIII and 
higher.}\\
\hline
Ar (2.9-3.1 {\rm  keV}) & 1.9 & 3.4 & 5.0 & 272$^b$\\
\hline 
Fe (6.4-6.9 {\rm  keV}) & 0.8 & 1.4 & 2.1 & 78$^b$\\
\hline
\hline
  $T^{(2)}$ [10$^7$ K] & 1.2 & 2.8 & 6.6 &\\
\hline
\end{tabular}
\end{minipage}
\end{table}

%Tab ******************************

\section{Discussion}

There are observational evidences for the clumpy structure of SN ejecta
in the remnants  of different types and ages.
Observations are available for the core-collapsed SNRs (e.g. Cas A)
and also for the remnants are thought to be of Type Ia (see e.g.
 the recent \xmm\ observations of Tycho by Decourchelle \etal (2001)).
Some of the SNRs are young (e.g. Cas A) and others are older (e.g.
the Vela SNR).

The observation of the ejecta clumps composition
would provide a valuable test to study the details of SN phenomena.
The simulations of SN nucleosynthesis (e.g. Woosley \& Weaver 1995;
Thielemann \etal, 1996) predict the production of $\sim 0.1 \Msun$ of
$^{56}$Ni and $^{28}$Si and $\gsim 0.003 \Msun$ of
$^{36}$Ar  in a core-collapsed SN. The metal distribution
through the ejecta at the postexplosion stage is still poorly known.
There are conclusive evidences for large-scale macroscopical mixing of Ni
in the SN 1987A ejecta (e.g. McCray 1993). However the modeling of the
$^{56}$Fe distribution through the ejecta at the microscopic level is
 rather a complicated task. This is because of the energy deposition
effect from $^{56}$Ni decay (see e.g. Wang \& Chevalier 2001;
Blondin, Borkowski \& Reynolds 2001 for recent discussions).
The X-ray line emission in our model is less sensitive to the details of
the element distribution in
a fragment because of the high penetrating ability of energetic
electrons responsible for the line excitation.
Even if the Fe (or other element) atoms are locked
in the dust grains we would still have X-ray line emission excited by
fast electrons.

We show that the X-ray line luminosities and spectra are
 different for the SN ejecta fragments in a dense molecular
cloud and for that in a tenuous medium. The efficiency
of X-ray line production is higher in the
fragments of lower ionization state. Strong MHD shocks could transform
a sizable fraction of the kinetic energy into nonthermal particles
thus reducing the heating of the shocked gas and increasing its compression.
The column density of the shocked gas
could reach $\sim 10^{21} \cms$ and the ion temperatures just behind
the shock transition region is below 5$\cdot 10^6$ K.
Thus, the fast fragments (of velocity $v_{\rm k} \lsim 1,500 \kms$)
moving through  molecular clouds of density $> 10^3 \cmc$ would drive
the {\it radiative} bow shock wave.
Hollenbach \& McKee (1989) have modeled the spectra of
the radiative shocks. They found rich spectra of H$_2$ emission
as well as atomic fine-structure lines with strong [OI] (63 $\mu m$) [OIII],
[Fe II], [C II] (158 $\mu m$), [Si II](35 $\mu m$), 
[Ne II] (12.8 $\mu m$) lines.
That implies a correlation of IR and
optical emission of the fragment with the X-ray line emission dominated by
the lines of relatively low ion charge states.
The soft thermal X-ray continuum emission from the postshock gas
of T $\lsim 10^7$ K is not expected to be observable in that case.
On the other hand the nonthermal bremsstrahlung emission with the
hard spectrum of a typical photon indexes  $<$ 1.5  is predicted
in our model.
%Fig. 2 shows the calculated continuum emission from the "standard"
%fragment of velocity  2,700 $\kms$ interacting with a molecular cloud.
The continuum bremsstrahlung emission is rather sensitive to the ionization
state of the fragment.
Our simulations show that the X-ray line luminosity is
 $\lsim 10^{-4}$ of the kinetic energy dissipation rate in the
fragment's bow shock.

The lifetime of the X-ray line emitting fragments in a dense molecular
cloud is expected to be a few times less than the fragment
deceleration time. Such a fragment should appear as
a nonthermal source showing a variability on a timescale of years.
The variability is most important for radio emission.
Synchrotron radio emission at a level about 100 {\rm mJy} {\rm arcsec}$^{-2}$
at 100 {\rm MHz} could be
expected from the fragments interacting with a dense molecular cloud if the
ambient magnetic field is about 0.1 {\rm mG}.
Radio emission from a fragment propagating through a tenuous medium is
much fainter. However, MHD type instabilities may greatly enhance the
magnetic
fields along the fragment boundary providing  substantial radio emission,
see Anderson \etal (1994); Jones \etal (1996).

The X-ray line luminosity
$L_{\rm x} \sim$ 10$^{31}~ \ergs$ (per 10$^{-4}~\Msun$ of Si, S, Ar, Ca, Fe)
is predicted from an individual SN fragment
in a molecular cloud. It could be detected from
a few {\rm kpc} distance with the current detectors aboard \chan\ and \xmm\ .
An obvious candidate to study is IC 443:  the SNR interacting with a molecular
cloud. The recent \xmm\ observation of IC 443 by Bocchino \& Bykov (2001)
has resolved the hard X-ray source 1SAX J0618.0+2227 correlated with
a molecular cloud into two sources one of which is generally consistent
with the ejecta fragment interpretation.
A dedicated observation of 1SAX J0618.0+2227 is desirable  to
check the interpretation.

An ensemble of unresolved
SN fragments could contribute substantially to the observed diffuse
iron line emission. Diffuse iron line emission has been found by {\it ASCA}
from the Galactic Centre region on a scale about a degree along the galactic
plane (Koyama \etal 1996) and from a more extended region of  
the Galactic Ridge (e.g. Tanaka 2002). 
He-like and H-like Fe-K lines analyzed by Tanaka (2002)
are significantly broadened corresponding to a velocity dispersion of a
few thousand $\kms$ . The high dispersion can be naturally explained if
the lines are due to unresolved faint SN fragments. Alternatively,
Tanaka (2002) has suggested that the high dispersion
is due to charge-exchange processes of low-energy cosmic rays
providing the high velocities of the line emitting ions.

A molecular cloud irradiated by a
hard X-ray source is expected to be observable
in fluorescent X-ray lines as
X-ray reflection nebulae -- a new class of X-ray sources
(e.g. Sunyaev \etal 1993; Sunyaev \& Churazov 1998).
The observed iron line emission from the Sgr B2 (Murakami \etal
2001) and Sgr C (Murakami \etal 2001a) complexes was considered by
the authors as the X-ray reflection nebulae irradiated by a bright
source in the GC which was active in the past.
Clumps of X-ray line  emission of neutral and highly ionized iron
of equivalent widths $\sim$ {\rm  keV} and absorption corrected 2-10 {\rm  keV}
fluxes $\sim 10^{-12} \enf$ was found by \chan\ in the Galactic center
region (Bamba \etal 2002). The photon indexes of 2-10 {\rm  keV} continuum
emission -- while not too constrained -- are broadly consistent with
the photon indexes $\leq$ 1.5 predicted by the model of SN fragments
emission considered above. If a substantial fraction of the iron
ejected by an SN in a molecular cloud would reside in the fast moving
fragments we could expect the iron line luminosity of
$L_{\rm x} \sim$ 10$^{34}~ \ergs$ per SN which is consistent with
that observed from the Sgr C complex.
The line emission from the Sgr B2 is somewhat brighter and would require 
contributions of more than one SNe.
Infrared line observations from radiative
shocks that accompany SN fragments in molecular clouds would help
to distinguish the SN fragments contributions.
A correlation between SiO (J=2$\rightarrow$0) emission morphology and
6.4 {\rm  keV} Fe line emission on the Galactic Center large scale  and
also within
Sgr A and B regions found recently by Martin-Pintado \etal (2000)
could be consistent with the multiple SN fragments contribution.
{\it ISO} observations of 18 molecular clouds in 
the Galactic Center  region were performed 
by Martin-Pintado \etal (2000a). Towards
most of the clouds they have detected the "ionized bubbles" with 
the fine structure line emission of ionized species: SIII, NeII, 
and in some cases also NeIII, NII, NIII, OIII. 
The "bubbles" could be relevant to the fragment-cloud interactions.

Nonthermal hard X-ray continuum emission from the Galactic ridge
 requires a 
population of accelerated electrons (e.g. Valinia \etal 2000,
Dogiel \etal 2002).
Hard continuum of the SN fragments 
%of photon indexes $\lsim$ 1.5 
could contribute to the observed emission. 
The galactic diffuse
emission mapping with forthcoming {\it INTEGRAL} mission will
address the issue.

\begin{acknowledgements}
I thank the referee for very constructive comments.
The work was supported by INTAS-ESA 99-1627 grant.

\end{acknowledgements}
%\appendix

%\section*{References}

%\begin{flushleft}
%\medskip

%\small

%\end{flushleft}
\end{document}